\documentclass[conference]{IEEEtran}
\IEEEoverridecommandlockouts
\usepackage{cite}
\usepackage{amsmath,amssymb,amsfonts,amsthm}
\usepackage{enumerate}
\usepackage{graphicx}
\usepackage{textcomp}
\usepackage{caption}
\usepackage[labelfont=bf]{caption}
\usepackage[table]{xcolor}
\usepackage[acronym]{glossaries}
\usepackage{float}
\usepackage{subcaption}
\usepackage{booktabs}
\usepackage{amsmath}
\usepackage{algorithm}
\usepackage{algpseudocode}
\usepackage{hyperref, verbatim}

\def\BibTeX{{\rm B\kern-.05em{\sc i\kern-.025em b}\kern-.08em
    T\kern-.1667em\lower.7ex\hbox{E}\kern-.125emX}}

\begin{document}
\include{glossary}

\title{Simulating Application Behavior for Network Monitoring and Security}

\author{\IEEEauthorblockN{Murugaraj Odiathevar, Kim Chung Yup
}
	\thanks{\hspace*{-\parindent}\footnotesize}
\IEEEauthorblockA{\textit{Research and Business Foundation}, 
\textit{Sungkyunkwan University} \\ \textit{2066, Seobu-Ro, Jangan-Gu, Suwon-Si, Gyeonggi-Do, Republic of Korea, 16419}\\ 
muru.raj@skku.edu, yup.kim@skku.edu}
}

\maketitle

\begin{abstract}
Existing network simulations often rely on simplistic models that send packets at random intervals, failing to capture the critical role of application-level behavior. This paper presents a statistical approach that extracts and models application behavior using probability density functions to generate realistic network simulations. By convolving learned application patterns, the framework produces dynamic, scalable traffic representations that closely mimic real-world networks. The method enables rigorous testing of network monitoring tools and anomaly detection systems by dynamically adjusting application behavior. It is lightweight, capable of running multiple emulated applications on a single machine, and scalable for analyzing large networks where real data collection is impractical. To encourage adoption and further testing, the \emph{full code is provided as open-source}, allowing researchers and practitioners to replicate and extend the framework for diverse network environments.

\end{abstract}


\section{Introduction}
Network simulation plays a crucial role in the design, analysis, and evaluation of computer networks by enabling researchers and engineers to study network behavior without requiring physical infrastructure. By providing a controlled environment for testing, simulation tools help assess network performance, optimize configurations, and detect potential issues before real-world deployment. Popular network simulators, such as Cisco Packet Tracer and NetCracker, offer functionalities for modeling network operations, defining traffic characteristics, and generating statistical reports, making them invaluable for both academic research and industry applications.

Despite these advantages, most network simulation tools primarily focus on modeling network-layer interactions rather than accurately replicating application-level behavior. Traditional simulation approaches typically generate synthetic traffic patterns based on packet-level characteristics without considering the higher-layer dynamics that arise from real application usage. This limitation reduces the realism of simulated network environments, particularly in the context of monitoring, anomaly detection, and security analysis.

To address this gap, this work proposes a fully statistical approach that models network behavior by first simulating application-layer interactions and deriving the corresponding network traffic from them. Unlike conventional methods that rely on predefined traffic models or require captured real-world data, this approach probabilistically generates network traffic based on application behavior, ensuring flexibility, scalability, and reproducibility. Furthermore, it does not require complex hardware setups, making it accessible for both large-scale simulations and lightweight testing environments.

The overarching goal of this work is to generate realistic packet captures (PCAPs) that mimic actual application network flows. By leveraging probability distributions to model application behavior, the generated traffic can be used for network monitoring, security analysis, and anomaly detection research. The following sections provide a detailed discussion of existing network simulation methodologies, their limitations, and how the proposed statistical framework advances the field. 

\section{Literature Review}

Network simulation is a widely used technique for evaluating communication protocols, security mechanisms, and performance characteristics in a controlled environment. It provides scalability and cost-effectiveness while eliminating the need for physical infrastructure. However, traditional network simulations often lack the realism observed in testbeds that employ real devices and applications \cite{gomez2023survey}. Various network simulation tools exist, each offering distinct capabilities, making selection and implementation a challenge for users \cite{smera2022networks}. While many frameworks focus on protocol-level interactions, they often overlook application-specific behavior, which plays a crucial role in real-world network performance.

Several simulation frameworks have been developed to bridge the gap between abstract network models and real-world deployments. For instance, the \textit{Network Simulation Bridge (NSB)} facilitates integration between real applications and network simulators, ensuring scalability while minimizing performance overhead \cite{kuttivelil2023network}. Similarly, the \textit{Network Research Simulator (NRS)} provides a flexible programming interface to conduct experiments on network robustness, including simulations of large-scale attacks and predictive analytics using AI \cite{marzo2022network}. Despite their contributions, these approaches do not fundamentally model application-level behavior but rather provide simulation support for network-layer functionalities.

Efforts have been made to incorporate real application traffic into network simulations. The use of captured application-specific traffic flows allows for the modeling of application performance under different network conditions \cite{zacks2021systems}. Studies on web traffic classification have demonstrated that application-level traffic exhibits distinct statistical properties that can be leveraged for simulation and anomaly detection \cite{karayaka2022application}. However, such approaches rely on empirical data collection rather than a fully statistical modeling approach, making them less adaptable to varying conditions. Furthermore, solutions like Android client emulation \cite{hetu2014similitude} provide a means to simulate user interactions but require resource-intensive setups.

A key limitation of existing methodologies is the lack of a true application-layer modeling approach. While application-level traffic data has been correlated with network-level traffic for improved insight \cite{sharma2021correlating}, most studies still treat applications as static sources of predefined traffic patterns rather than dynamic entities that influence network behavior. Probabilistic models generated from historical network traffic have been proposed to emulate application statistics on virtualized environments \cite{ganapathi2020data}, but these models do not fully replicate the underlying application behavior. Similarly, simulation tools for enterprise multicast \cite{wu2022ip} focus on protocol efficiency rather than the behavioral intricacies of applications.

The challenge of integrating real application behavior into network simulations has been long recognized. Existing simulators like NS3 offer extensive functionality for protocol testing, but their credibility is often questioned due to the lack of real-world application interactions \cite{rampfl2013network}. Other approaches attempt to introduce real applications into simulated networks \cite{ meszaros2019inet}, but they require hardware-intensive setups and are not always easily replicable. While existing network simulation techniques provide insights into network operations, they do not fundamentally simulate how applications influence network behavior \cite{kuttivelil2023network}.

Given these limitations, this paper proposes \emph{a fully statistical approach that replicates application behavior and builds network behavior from it}. Unlike traditional methods, which either simulate network traffic without understanding application behavior or require empirical traffic captures, this approach \emph{models application behavior probabilistically using statistical distributions}, allowing it to dynamically generate network traffic patterns. This methodology is \emph{easily replicable}, as it does not depend on complex data collection processes or specialized hardware. Additionally, the statistical nature of the approach makes it \emph{lightweight and adaptable}, enabling its use across different network conditions without significant overhead. By focusing on application-layer behavior as the foundation of network simulation, this work advances the realism and usability of network simulation tools in monitoring, anomaly detection, and performance analysis.

\section{Methodology}
PCAP (Packet Capture) files are data files used to record and store network packet data captured over a network. They are generated by packet sniffing tools such as Wireshark, tcpdump, or other network monitoring utilities. These files provide a detailed snapshot of network traffic, capturing individual packets along with their metadata, such as source and destination IP addresses, port numbers, timestamps, and payload data and for network analysis, troubleshooting, and security purposes, such as identifying network issues, analyzing communication protocols, and detecting malicious activities. Typical PCAP analysis includes examining packet headers, identifying protocols in use (e.g., TCP, UDP, HTTP), and reconstructing data flows to understand application behavior or detect anomalies. Due to their detailed nature, PCAP files are essential for network engineers, cybersecurity experts, and researchers. 
Tools such as Scapy \cite{rohith2018scapy} library in Python provides a powerful framework for packet crafting, network traffic analysis, testing and with other libraries it can be used to simulate application behaviour. Other tools include the libpcap (Linux/Unix) \cite{garcia2008programming}, WinPcap (Windows)\cite{xiaoguang2016packet} libraries.

The first step is to monitor the interface to extract pcap data and determine the application behaviour over a period. This can be a few minutes, an hour, a day or a week based on the available memory and patterns. Most applications differ in patterns in weekday vs weekends or holidays, daytime vs night time as they depend on users. It also depends on how much detail is required in modeling the network. Capturing changes by the minute or even hour in most cases is not necessary. Most purposes of network monitoring are to detect when the network can be overloaded through a missed router configuration or a malfunctioning application, and the focus of this paper is on the latter. Applications exhibit distinct network traffic patterns based on their functionalities and user interactions. Understanding these patterns is crucial for effective network management and security. Most applications use the traditional client-server communication (Request-Response). Other types include Push-Based Communications \cite{c2019overview}, Pub-Sub model\cite{kermarrec2013xl}, fire and forget\cite{subramaniam2006fire} or broadcast models \cite{braverman2015information}. 

\begin{figure}[!htbp]
    \centering
    \caption{Distribution of methods (Example)}
    {
        \includegraphics[width=0.4\textwidth]{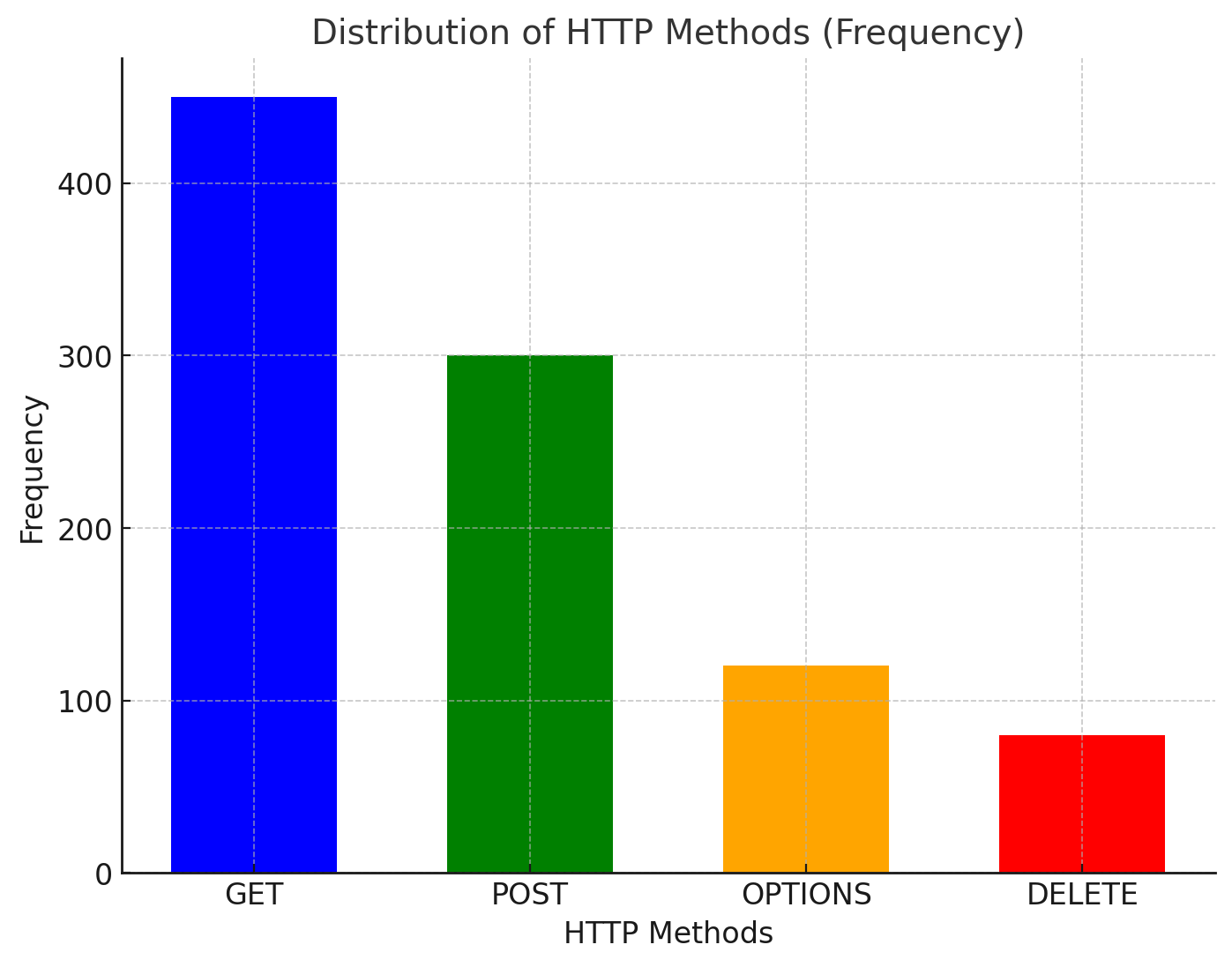}
        \label{fig:freq_meth}
    }
    \vfill
    {
        \includegraphics[width=0.30\textheight]{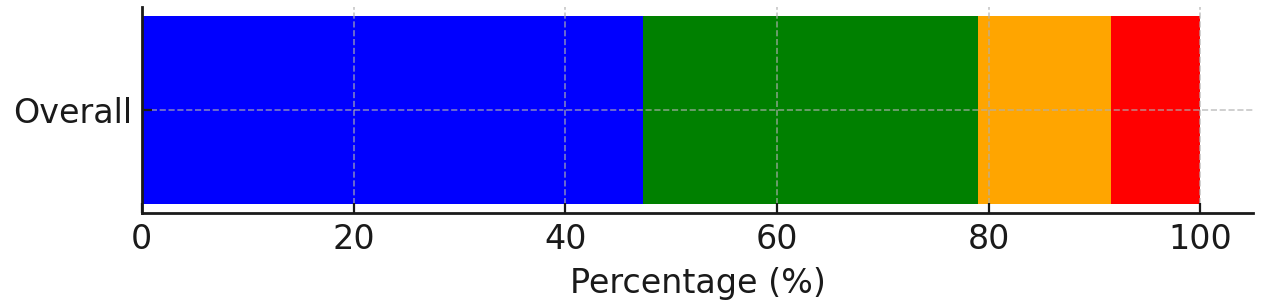}
        \label{fig:pct_meth}
    }
    \label{fig:distribution_methods}
\end{figure}

Regardless of the method of communication, each application will have a few methods. In the example of ``Request-Response" type communication - 'GET', 'POST', 'OPTIONS', 'DELETE'. Aggregating the type for each application, over a period of time, a histogram can be determined. An example of HTTP methods distribution are shown in figure \ref{fig:distribution_methods}. These proportion can be easily obtained after aggregating the pcap files for each application over a period of time. The distribution can then be modeled using a Uniform distribution.

The next step is to determine the volume of data or the payload transmitted over the network for each type of connection for each application. The distribution of the payload can vary in both size and shape. For example, web traffic in http can follow heavy-tailed distribution such as Pareto distribution, file downloads may follow log-normal distribution and VOIP can follow exponential distribution. These patterns have been used for traffic classification \cite{finsterbusch2013survey,zhang2013unsupervised}. 

\begin{figure}[!htbp]
    \centering
    \caption{Distribution of Payloads (Example)}
    
    \subfloat[$a=50, b=1450, \alpha=2, \beta=5$]
    {\includegraphics[width=0.45\columnwidth]{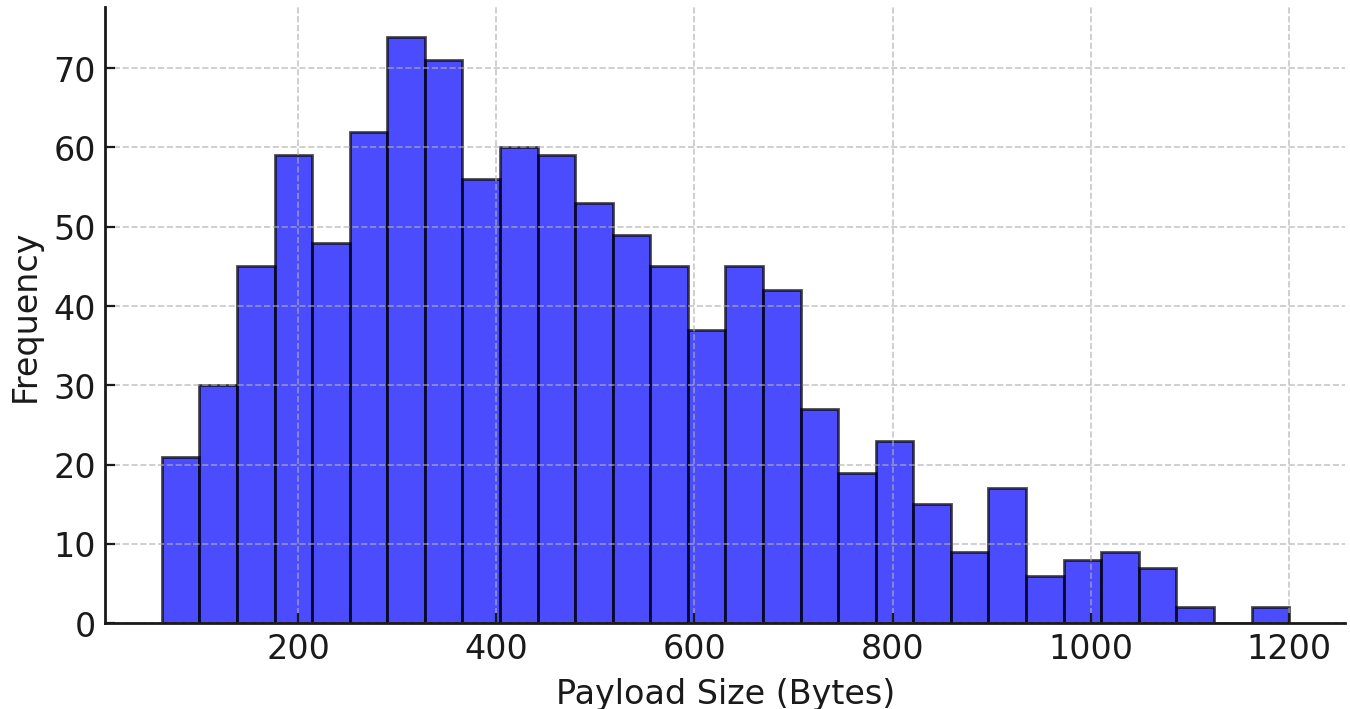}%
    \label{fig:pd1}}
    \hfill
    \subfloat[$a=50, b=1450, \alpha=5, \beta=2$]
    {\includegraphics[width=0.45\columnwidth]{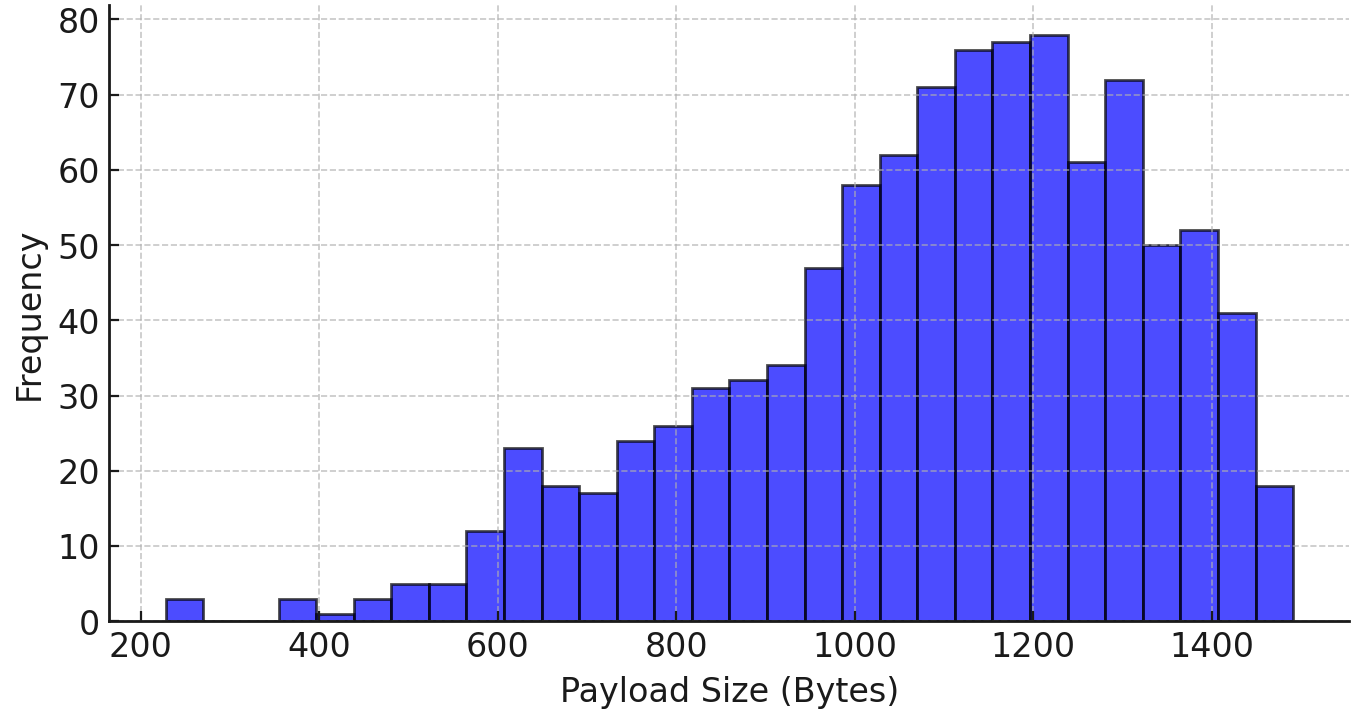}%
    \label{fig:pd2}}
    
    \vspace{1em} 
    
    \subfloat[$a=50, b=1450, \alpha=2, \beta=2$]
    {\includegraphics[width=0.45\columnwidth]{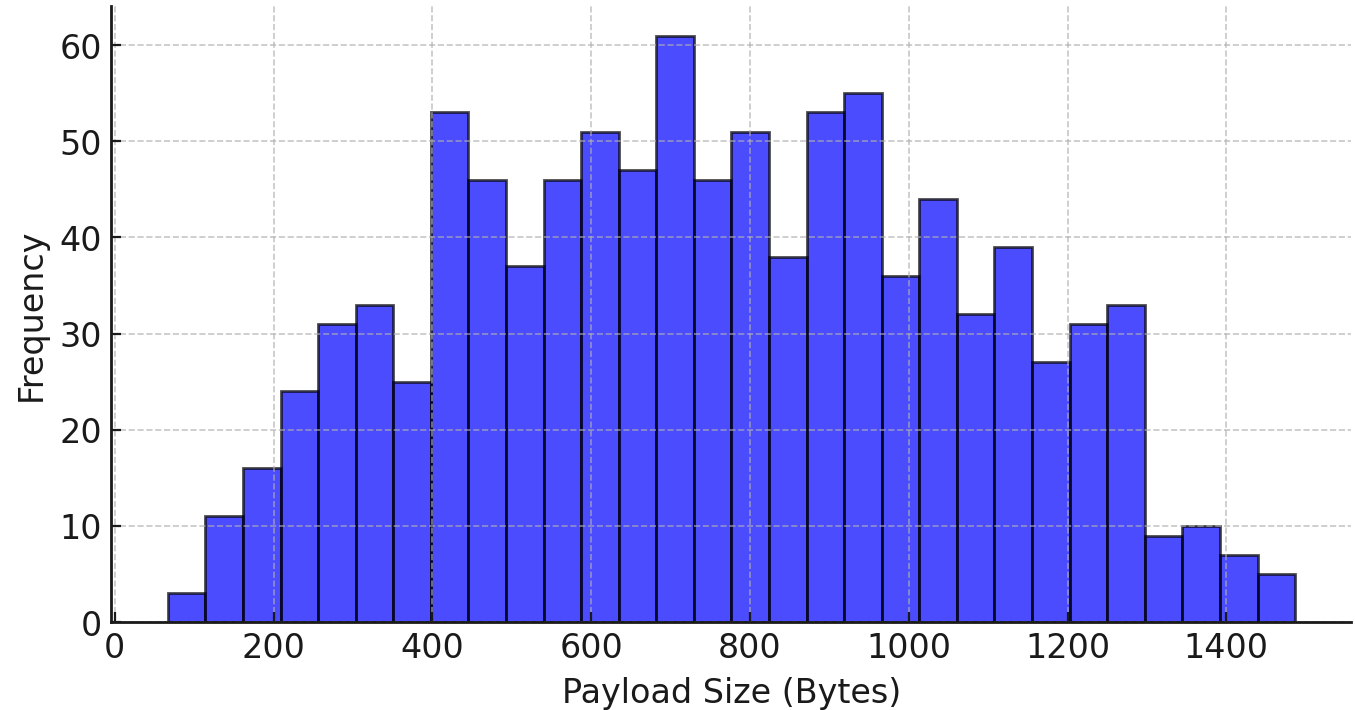}%
    \label{fig:pd3}}
    \hfill
    \subfloat[4: $a=50, b=1450, \alpha=0.5, \beta=0.5$]
    {\includegraphics[width=0.45\columnwidth]{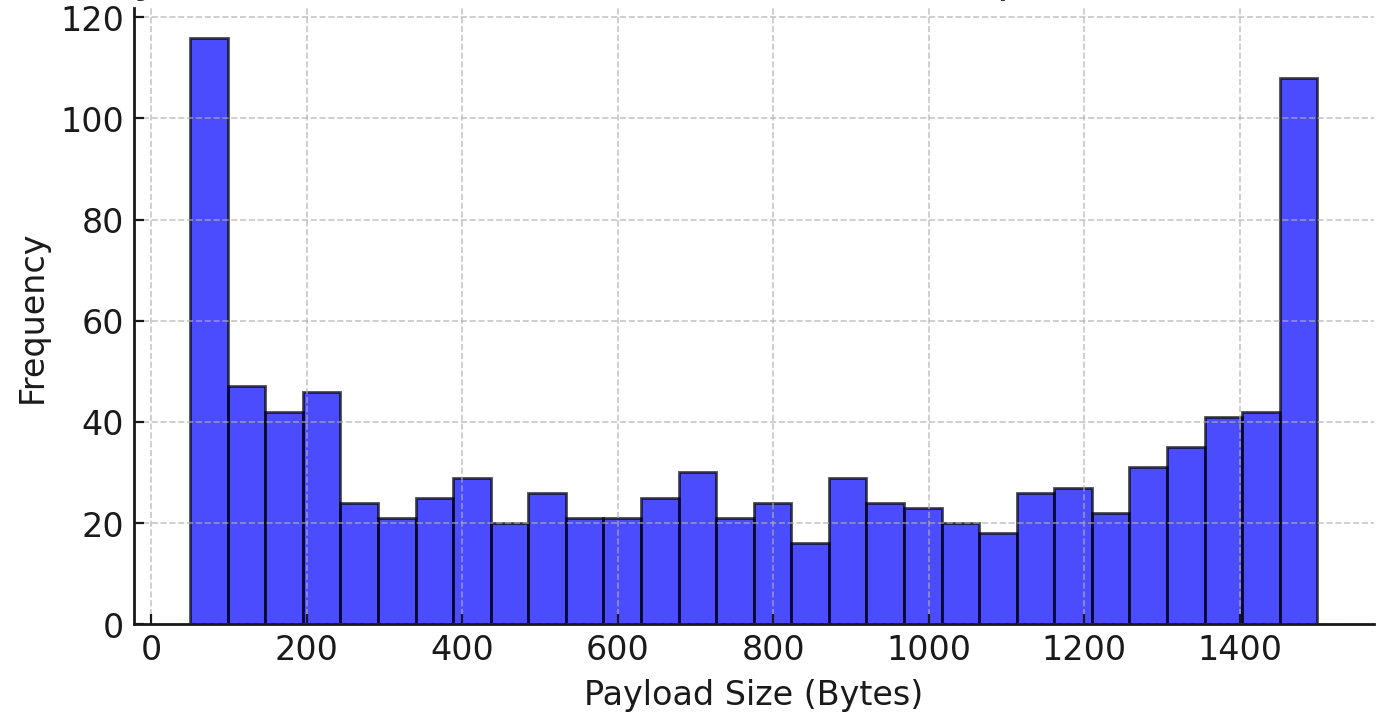}%
    \label{fig:pd4}}
    
    \label{fig:pdall}
\end{figure}

For the simulation, the absolute payload sizes can be modeled using a scaled and shifted Beta distribution. Equation \ref{eq:betadist} contains four changeable parameters to provide flexibility for different types of absolute payload sizes for each packet. Once the shape of the payload is determined after observing the data, the parameters $a$, $b$, $\alpha$ and $\beta$ can be determined. Figures \ref{fig:pdall} give four possible shapes for the payload sizes and they can be scaled using the $a$ - minimum payload size observed and $a+b$ - maximum payload size observed. The advantages of using Beta distribution are as follows. (i) can capture skewness (small packets are common, large packets are rare) and variability. (ii) It retains the flexible shape characteristics of the Beta distribution while allowing the data range to shift and scale. (iii)Ensures the output stays within the specified range [a, a+b]. (iv) Many real-world network metrics are bounded and can be effectively modeled using this transformation \cite{gupta2020fundamentals}.

\begin{equation}\label{eq:betadist}
    X \sim a + b \cdot \text{Beta}(\alpha, \beta)
\end{equation}

The next important factor to consider is the maximum number of connections which is determined by the maximum number of users for each application over a period of time. This number can be determined based on the number of unique IP addresses in the data. With these information, a configuration file for each application can be developed as shown in table \ref{app_stat_config}. The ``Type" in the table is for HTTP requests and a few other TCP packets. These categories are changeable depending on the applications and their protocols. 

\begin{table}[h!]
    \centering
    \caption{Method Distribution and Connections}
    \label{tab:method_distribution}
    \begin{tabular}{@{}lcc@{}}
        \toprule
        \textbf{Type, M} & \textbf{Probability} \\ 
        \midrule
        GET       & 0.52    \\
        POST      & 0.12   \\
        OPTIONS   & 0.23   \\
        TCP       & 0.02    \\
        TEARDOWN  & 0.11  \\
        \midrule
        Total Connections, N & 25  \\
        \midrule
        a & 1  \\
        b & 739 \\
        $\alpha$ & 5  \\
        $\beta$ & 24  \\
        \bottomrule
    \end{tabular}
\end{table}\label{app_stat_config}

The distribution of the response codes [200, 304, 302, 401, 400, 500, 404, 303] can be modeled using uniform distribution after aggregating the data in a similar fashion to \ref{fig:distribution_methods}, $W_{response\_code}$. A list of server ips addresses can be randomly determined for the applications and another much longer list of ip addresses can be generated for client ips. Server port numbers can also be defined likewise. Here is the algorithm.

\begin{algorithm}[!htbp] 
\caption{Simulate Network Traffic and Generate PCAP}
\begin{algorithmic}[1]
\Require $\text{server\_ips}, \text{client\_ips}, \text{app\_config}, \text{server\_ports}, T_{\text{time}}$
\Ensure $\text{output\_pcap}$

\State Initialize $\text{response\_codes}$
\State Compute $\text{response\_code\_pool}$ using $W_{\text{response\_code}}$
\State $T \gets 0$

\While{$T < T_{\text{time}}$}
    \State Randomly select $k$ applications $\{A_{i_1}, \ldots, A_{i_k}\}, k = 50$
    \For{application $A_i$}
        \State Retrieve $P_{\text{type}}, N_{\text{connections}}$ from $\text{app\_traffic\_config}$
        \State Generate inter-arrival times $\Delta t \sim \text{Exponential}(\lambda)$
        \For{$j \gets 1$ to $N_{\text{connections}}$}
            \State Select $M_j \sim P_{\text{type}}$
            \State Randomly assign $\text{client\_ip}, \text{server\_ip}, \text{server\_port}$
            \State Generate packets:
            \State \hspace{1em} \textbf{Handshake:} SYN, SYN-ACK, ACK
            \State \hspace{1em} \textbf{Data Exchange:} Response code
            \State \hspace{1em} \textbf{Payload:} Simulate $M_j$ using $X$
            \State \hspace{1em} \textbf{Teardown:} FIN, FIN-ACK, ACK
            \State Update $T \gets T + \Delta t$
        \EndFor
    \EndFor
    \State Adjust timestamps to base time $T_{\text{base}}$
\EndWhile

\State Sort $\text{all\_packets}$ by timestamps
\State Write packets to PCAP: $\text{wrpcap}(\text{output\_pcap}, \text{all\_packets})$
\Return $\text{output\_pcap}$
\end{algorithmic}
\end{algorithm}

In our simulation, a 100 application configurations were defined of which 50 applications were randomly selected in each loop. For each type of connection $M$, a separate function is defined to generate a payload using the Beta distribution. The algorithm can be further refined to generate duplicate packets with r\% probability. In our network duplicate packets were detected around 6\% of the time. A long session function can be added on top of the methods to generate multiple packets from the same connection $M$. The simulation runs for a duration $T_{time}$ but due to the heavy processing and lower computational resources, the code might take longer to run which results in the simulation for the applications starting one after another. Though parallel processing would help, it is still challenging to simulate 50 processes at once. Hence, the time stamp is adjusted for application data in that loop to begin at $T_{base}$; when the first simulation in that ``for-loop" began.     

Once the simulation is complete, the pcap can be aggregated using Netflow methodologies \cite{hofstede2014flow} and the network statistics can be gathered. Features such as number of concurrent flows per client or server, number of response codes in categories of 200, 400 or 500, packet per second, bytes per second, number of request packets, number of response packets and many more can be determined to analyse the network pattern. These features can be used to build machine learning models for network monitoring and anomaly detection.

\section{Conclusion and Future work}

To conclude, the advantages of the simulation methodology are as follows.
\begin{enumerate}
    \item Realistic Variation: Through random distributions (exponential inter-arrivals, Beta-based payload sizes), random packet duplication, and method-based flows, the simulation yields complex and lifelike PCAPs.
    \item Application Profiles: The configuration dictionary ensures each simulated application can have unique traffic characteristics.
    \item Extensibility: Additional methods (like PUT, HEAD, or custom protocols) can be added with minimal changes. The approach is modular enough to integrate new ideas.
    \item Easy to implement: Through pure statistical distribution and modifying pcap timestamps, large network data can be simulated on relatively small hardware.
\end{enumerate}

The above algorithm implements simulations for TCP connections and it  can be complimented with other protocols in future work. Furthermore, the parameters $a$, $b$, $\alpha$ and $\beta$ and the weights for the uniform distribution can also be determined using machine learning. The simulation can be tested with live network data. Code available on github for open source testing: \href{https://github.com/muru-raj10/AppBehaviourNetworkSim}{https://github.com/muru-raj10/AppBehaviourNetworkSim}

\bibliographystyle{IEEEtran}
\bibliography{main_en}
\end{document}